\documentclass[fleqn,usenatbib]{mnras}

\usepackage{newtxtext,newtxmath}

\usepackage[T1]{fontenc}
\usepackage{ae,aecompl}

\usepackage{graphicx}	
\usepackage{amsmath}	
\usepackage{amssymb}	

\usepackage[normalem]{ulem}

\newcommand {\0}   {\phantom{0}}
\newcommand {\da}  {\ensuremath{\Delta a_0}}
\newcommand {\dr}  {\ensuremath{\Delta r}}
\newcommand {\fma} {\ensuremath{f_M}}

\title[Orionid outbursts observed by the Maya]{Orbital dynamics of highly
probable but rare Orionid outbursts possibly observed by the ancient Maya}

\author[J. H. Kinsman and D. J. Asher]{
J. H. Kinsman$^{1}$\thanks{E-mail: jhkinsman@gatech.edu}
and D. J. Asher$^{2}$\thanks{E-mail: David.Asher@armagh.ac.uk}
\\
$^{1}$P.O. Box 723, Murrayville, Georgia, 30564, USA\\
$^{2}$Armagh Observatory \& Planetarium, College Hill, Armagh BT61 9DG, UK
}

\date{Published as MNRAS \textbf{493,} 551--558 (2020). Accepted 2020 January 17. Received 2020 January 3; in original form 2019 May 30}

\pubyear{2019}

\begin{document}
\label{firstpage}
\pagerange{\pageref{firstpage}--\pageref{lastpage}}
\maketitle

\begin{abstract}
\\
Using orbital integrations of particles ejected from Comet Halley's passages
between 1404 BC and 240 BC, the authors investigate possible outbursts of the
Orionids (twin shower of the Eta Aquariids) that may have been observed in
the western hemisphere. In an earlier orbital integration study
the authors determined there was a high probability linking probable outbursts of the Eta Aquariid meteor shower with certain events recorded in inscriptions during the Maya Classic Period, AD 250--900. 
This prior examination was the first scientific inquiry of its kind into
ancient meteor outbursts possibly recorded in the western hemisphere where
previously no pre-Columbian observations had existed. In the current paper
the aim is to describe orbital dynamics of rare but probable Orionid
outbursts that would have occurred on or near applicable dates recorded in
the Classic Maya inscriptions.
Specifically,
significant probable outbursts are found in AD 417 and 585 out of 30 possible
target years.  The driving mechanisms for outbursts in those two years are Jovian 1:6 and 1:7 mean motion resonances acting to maintain compact structures within the Orionid stream for over 1~kyr.  Furthermore,
an Orionid outburst in AD 585 recorded by China is confirmed.

\end{abstract}



\section{Introduction: Halley's Comet and Orionids}
\label{Sintro}

Comet 1P/Halley's meteoroid stream produces twin showers: the Orionids (IAU
meteor shower 008 ORI) at the ascending node and the Eta Aquariids (031 ETA)
at the descending node.  The authors' previous investigations
\citep{ka17} focused on the Eta Aquariid outbursts because the heliocentric
distance of the descending node of the parent comet was
approximately equal to the Earth's orbital radius near the
middle of the Maya Classic Period (i.e., $\sim$AD 530). In contrast,
the present paper's investigations focus on the Orionids in
spite of the fact that the ascending node of Halley was nowhere
near Earth's orbital radius by the Maya Classic time (AD
250--900).

Orionid outbursts during that epoch were
likely rather rare as exhibited by the historical record, where China was the
only ancient culture to record Orionid outbursts earlier than AD 900, and those were in AD 288
and AD 585. Similarly, in the present paper, the authors found only two probable
outbursts out of a database from the Maya corpus of
inscriptions\footnote{Of all the ancient cultures found in the
  western hemisphere, the Maya civilization is the only one with a rich
  corpus of dates \citep[cf.][sec.\ 1.1]{ka17} available for
  investigation. As stated in the authors' previous study \citep{ka17},
  currently there is no known definitive statement in the Maya corpus of a
  `shooting star' or meteor shower. The authors rely on the most recent
  correlation literature for converting Maya calendar dates into the Julian
  dates of the Christian calendar, the Martin-Skidmore correlation constant
  \citep{ms12}, 
  and the most recent carbon dating \citep{kennettetal13}.} that currently contains 30 different viable
`target years' (Sections \ref{Smeth} and \ref{Sres}) prior to AD 900. With such a small
sample size, the authors' goal is not to correlate events in the Maya record
to these two probable outbursts as in their previous study, but to analyze
the orbital dynamics of the outbursts themselves. The year 585 was not only
recorded in the Chinese annals, but remarkably was also found as one of the two
years in which a probable Orionid outburst could have been observed by the
Maya.

By the present epoch, the general tendency for particles
released $\sim$3 kyr ago is for the Orionid node to have precessed well
beyond Earth's orbit \citep[][fig.\ 6]{ryabova03}. \citet{sw07}, however,
demonstrated that there are observable Orionid outbursts even in the present
epoch resulting from particles released 3 kyr ago and trapped in a Jovian 1:6
resonance, raising the possibility that owing to a similar dynamical effect
Orionid outbursts could have occurred in the first millennium AD.

\begin{figure}
\includegraphics[width=\columnwidth]{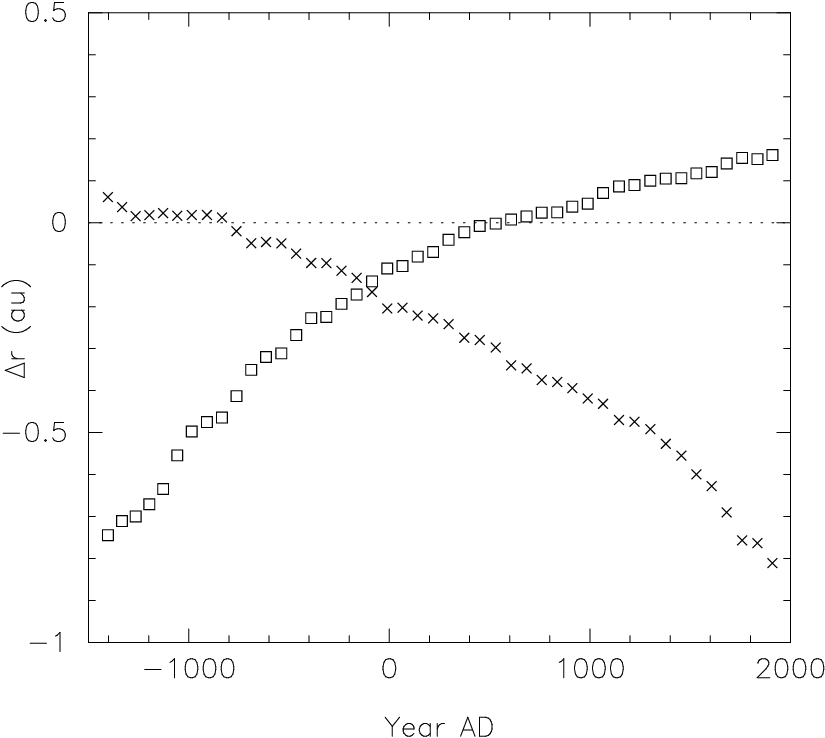}
\caption{Distance of Earth orbit from comet nodes at perihelion returns of
  1P/Halley listed by \citet{yk81}. Squares: descending node, corresponding
  to ETA. Crosses: ascending (ORI). Positive/negative \dr\ is Earth orbit
  outside/inside comet orbit.}
\label{Fdr}
\end{figure}

Thus given those present epoch Orionid outbursts, it is remarkable that the last time Halley's ascending node was equal to the Earth's orbital radius was approximately 800 BC (Fig.~\ref{Fdr}).
Thereafter the node continued to precess further outside Earth's orbit.  For a number of centuries prior to 800 BC, however, the ascending nodal distance maintained a constant value around 0.98 au from about 1266 BC until about 836 BC, sustaining very little precession.
Figure \ref{Fdr} shows that for particles ejected around 1000 BC the
precession rate must be slowed almost to nothing if Orionid intersections are
to occur during either the Maya Classic Period or the present epoch, whereas
the ascending node of Halley itself has moved a few $\times$ 0.1 au
outside Earth in the Classic and several $\times$ 0.1 au outside at present.
The precession rate of angular orbital elements of meteoroid stream
particles can be dramatically affected by resonances
\citep{fs86,sf88,ff92,ac93}.
In Section \ref{Smmr} we shall see that Orionid outbursts occur -- i.e.,
ascending nodes are Earth intersecting -- during the Maya Classic Period due
to particles trapped in resonance since release from Halley at its 1266 BC
and 911 BC returns.
The large tightly clustered segments of particles in Jovian resonances
actually impeded the precession of the meteoroid trails.

\citet{sa14} verified historical outbursts in the AD 1400's and 1900's and reported that Halley was trapped in a 1:6 resonance with Jupiter from 1404 BC to 690 BC. The two conditions of Halley's ascending node being very close to Earth and Halley being in a strong resonance with Jupiter for over seven centuries set up a unique situation:  particles released from Halley over this long 700 year period could result in particles trapped in resonance that would later impact Earth.    Amazingly, the result would be sharp displays of meteor outbursts due to particles released from the comet over a thousand years earlier.  Realizing the potential for outbursts during the Maya Classic Period due to resonant trappings added further impetus to investigate Orionid outbursts.

This paper will use orbital dynamics to
post-predict accurate dates and times of Orionid
outbursts that would have been observable by the ancient Maya civilization, potentially noted by 
dates recorded in extant inscriptions.
Computations using the same dynamical technique successfully
match the ancient Chinese records from AD 585. \\\\

\section{Methodology}
\label{Smeth}

Our methodology for investigating Orionid outbursts was similar to that used
with the Eta Aquariid investigations: given a beginning epoch where particles
were ejected, we considered target years found in the Maya corpus where
those particles would produce a meteor outburst.
The authors considered a `target' year as a year found in the corpus of inscriptions that included a date that fell within a range approximately one week before to approximately two weeks following a typical Orionid shower, with `typical' taken as spanning a J2000 solar longitude range of 200--208\degr.

Although over 300 Maya sites (found in the present countries of Mexico,
Guatemala, El Salvador, Belize and Honduras) exist with hieroglyphic
inscriptions \citep{prager14}, only about 150 of these sites contain legible
dates \citep{mathews14}. Numbering well over 1000 dates, approximately 35 of
these fell within our criteria. Given that some of the events related to
these dates were unknown or of questionable decipherment, we chose 30 of the
remaining candidates from the Mathews database and additional sources
for the following entries in Table \ref{TAevents}:
1 \citep[p.120--121, plate 33]{sm86};
2,3 \citep[dates not certain]{mathews14};
4 \citep[plate 2,3]{adams99};
10,17,18 \citep{sc17};
11 \citep[p.\ II-38]{gm04};
13 \citep[p.\ II-36]{gm04};
14 \citep{ha16};
15 \citep[p.\ II-38]{gm04};
20 \citep{stuart15};
24 \citep{martin15};
26 \citep[p.325]{sf90};
27 \citep{vailhern18};
28 \citep[noted date is one of three options provided by Stuart]{stuart04};
29 \citep[p.\ II-91]{gm04};
other entries in Table \ref{TAevents} are from Mathews database.

Due to variations in initial orbital period and planetary perturbations, there was little chance of the particles staying together after several revolutions
\citep{plavec56,plavec57}. Therefore, the authors looked for particles that had been trapped
in clusters by resonant actions with the larger planets.
As the dynamical evolution of particles in a trail released at a given
perihelion return of the parent comet depends primarily on the particles'
orbital period \citep{plavec56}, the `dust trail model' comprises ejection in
tangential directions to the comet's trajectory at perihelion.
As before, the
authors adopted the difference,
$\Delta{a_0}$,
between the semi-major axes at ejection time of the particle, $a_0$ 
and of the comet,
$\mathit{a_c}$
to parametrize the trail:
$$\Delta{a_0} = a_0 - \mathit{a_c}$$


The solution values of \da\ are values where particles reach the node in the
target year, close to the target date in that year.
The forward integration from the ejection epoch yields quantities \fma, \dr\
and nodal longitude \textit{in the target year}, corresponding to given \da.
The density of the outburst in this dust trail model
is determined by a product of terms as follows
\citep[cf.][p. 18, eq.~5]{asher00}:
$$\mathit{f}_\mathit{r}(\mathit{r}_E - \mathit{r}_A) \mathit{f}_\mathit{a}(\Delta\mathit{a}_0) \fma $$
where $\mathit{f_r}$ is a function of $\Delta{r}$, the difference between the
heliocentric distances of Earth and the particle's ascending node;
$\mathit{f_a}$ is a function of $\Delta{a_0}$;
and $\mathit{f_M}$, the $\textit{mean anomaly factor}$
is proportional to the along trail spatial density of particles,
measuring how much the particles in the stream stretch out over time.
The highest density, i.e., the greatest ZHR occurs when $\Delta{r}$ is the smallest
(with some activity enhancement for $|\dr|$ up to a few $\times$ 0.001 au),
$\Delta{a_0}$ is the smallest, and $\mathit{f_M}$ is the largest
(typically \fma\ closest to 1.0 is the highest).
Whether or not the meteors are visible to the Maya astronomers depends on the time of day that Earth passes through the ascending nodal longitude of the particles.

A range of 16--22 au in $a_0$ covers the ejection speeds of particles
producing visual meteors, allowing also for the typical change in orbital
period due to radiation pressure \citep{ka17}. Following similar procedures
in the previous analysis, integrations were carried out beginning with 601
particles spaced 0.01 au apart, and further integrations were performed until
the particles converged on solution values \da, $\Delta{r}$ and $\mathit{f_M}$ 
showing an outburst at a specific time and date \citep{ka17}.

As in \citet{ka17}, orbits for 1P/Halley's perihelion returns were from
\citet[][table 4, orbit well known back to 1404 BC]{yk81},
initial state vectors of eight planets (planetary barycentres) were from JPL
Horizons \citep{giorgini96}, and the \textsc{radau} algorithm
\citep{everhart85} was used in the \textsc{mercury} integrator
\citep{chambers99}.

\section{Results}
\label{Sres}

\begin{table*}
\centering
\caption{A strong Orionid outburst likely occurred on AD 417 September 24 due
to meteoroid-sized particles from 1P/Halley in 911 BC trapped in 1:7
resonance with Jupiter.  First two columns: perihelion return of comet when
particles released into trail; semi-major axis difference from comet at that
time.  Remaining columns are trail's Earth encounter in 417: solar longitude;
date (Julian calendar); time (UT); trail centre's nominal miss distance; \fma\
(stretching factor along trail); visibility from Maya longitudes. }
\label{T417}
\begin{tabular}{cccccccc}
\hline
Trail & \da\ (au) & $\lambda_{\sun}$ (J2000)
                             &  Date  & Time  & \dr\ (au)  & \fma     & Vis \\
\hline
911 BC & +1.183 & 204\fdg057 & Sep 24 & 06:36
& +0.00110 & $-$0.016 & yes \\
911 BC & +1.175 & 204\fdg094 & Sep 24 & 07:30 & $-$0.00102 & $-$0.146 & yes \\
911 BC & +1.180 & 204\fdg100 & Sep 24 & 07:37 & $-$0.00116 &   +0.392 & yes \\
\hline
\end{tabular}
\end{table*}

\begin{table*}
\centering
\caption{Orionid outbursts in AD 585 September resulting from the 1P/Halley
911 BC (1:7 resonance) and 1266 BC trails (1:6 resonance), recorded by China
and possibly observed by the Maya.  $+$5m in the Vis column indicates the outburst occurred 5 minutes after the end of the referenced visual observation time, i.e. 25 minutes before sunrise, and $-5$m indicates the outburst occurred 5 minutes prior to radiant rise. For columns cf.\ Table \ref{T417}.}
\label{T585}
\begin{tabular}{ccccccccc}
\hline
Trail & \da\ (au) & $\lambda_{\sun}$ (J2000)
                             &  Date  & Time  & \dr\ (au)  & \fma     & Vis \\
\hline
1266 BC & +0.768& 202\fdg048& Sep 23& 08:13&   +0.00296&   +0.048& vis & Maya \\
1266 BC & +0.772& 202\fdg131& Sep 23& 10:14&   +0.00145& $-$0.388& vis & Maya \\
1266 BC & +0.773& 202\fdg145& Sep 23& 10:33&   +0.00120&   +0.554& vis & Maya \\
1266 BC & +0.775& 202\fdg178& Sep 23& 11:21&   +0.00057& $-$0.031& +5m & Maya \\
1266 BC & +0.781& 202\fdg311& Sep 23& 14:34& $-$0.00271& $-$0.043& $-$5m & China\\
1266 BC & +0.782& 202\fdg334& Sep 23& 15:07& $-$0.00315&   +0.054& vis & China\\
\0911 BC& +1.447& 203\fdg923& Sep 25& 05:27& $-$0.00168& $-$0.013& vis & Maya \\
\0911 BC& +1.343& 204\fdg134& Sep 25& 10:33&   +0.00105& $-$0.093& vis & Maya \\
\hline
\end{tabular}
\end{table*}

Integrations were performed for all Halley returns in the
1404 BC to 240 BC range for 30 target years (Table~\ref{TAevents}) during the Classic Period to find what particles reach the ascending node around ORI time in those years.  If those particles' heliocentric distance is near Earth's orbit, they can potentially produce a meteor outburst.

The authors found productive returns\footnote{ Additional weak returns were
  found for the years 630, 634, 648 and 714, however any activity would
  likely have produced only a slightly enhanced display of the shower if at
  all and not a quantifiable outburst.}  for two separate years, AD 417 and
585.  In 417 two segments of the same group of particles, trapped in a 1:7
resonance with Jupiter (Sec.~\ref{Smmr}),
likely impacted Earth in rapid succession early in
the morning on September 24, likely producing strong meteor displays.
Table~\ref{T417} shows the details of the impact of the particles on that
date.

Both outbursts, peaking within just a few minutes of each other and thus
overlapping, were relatively strong with robust mean anomaly factor \fma\ and
fairly close to a central impact with the difference between Earth's orbital
radius and the particle stream within $\sim10^{-3}$ au. The particles were
likely medium size in terms of visual meteors, since $\da \approx +1$ au
\citep[higher and lower \da\ corresponding to smaller and larger meteoroids
  respectively;][]{ma99}. Viewing conditions were optimum as the radiant at
the computed peak time was about 46\degr\ above the eastern horizon and the
waning crescent moon, 27.2 days old, was in Virgo and had not risen yet.
Five days following the probable outburst on September
24 the birth of an unknown ruler was recorded on
417 September 29.

In addition to the results from 417, the authors found that strong Orionid
outbursts occurred in 585 on September 23 due to a resonant cluster of
particles.  The first segment would have been visible to the Maya and the latter of that same cluster was visible to the Chinese later that same day
(Table~\ref{T585}). These meteoroid-sized particles were from the 1266 BC Halley
passage, trapped in a 1:6 resonance with the planet Jupiter
(Sec.~\ref{Smmr}).

\begin{figure}

\includegraphics[width=\columnwidth]{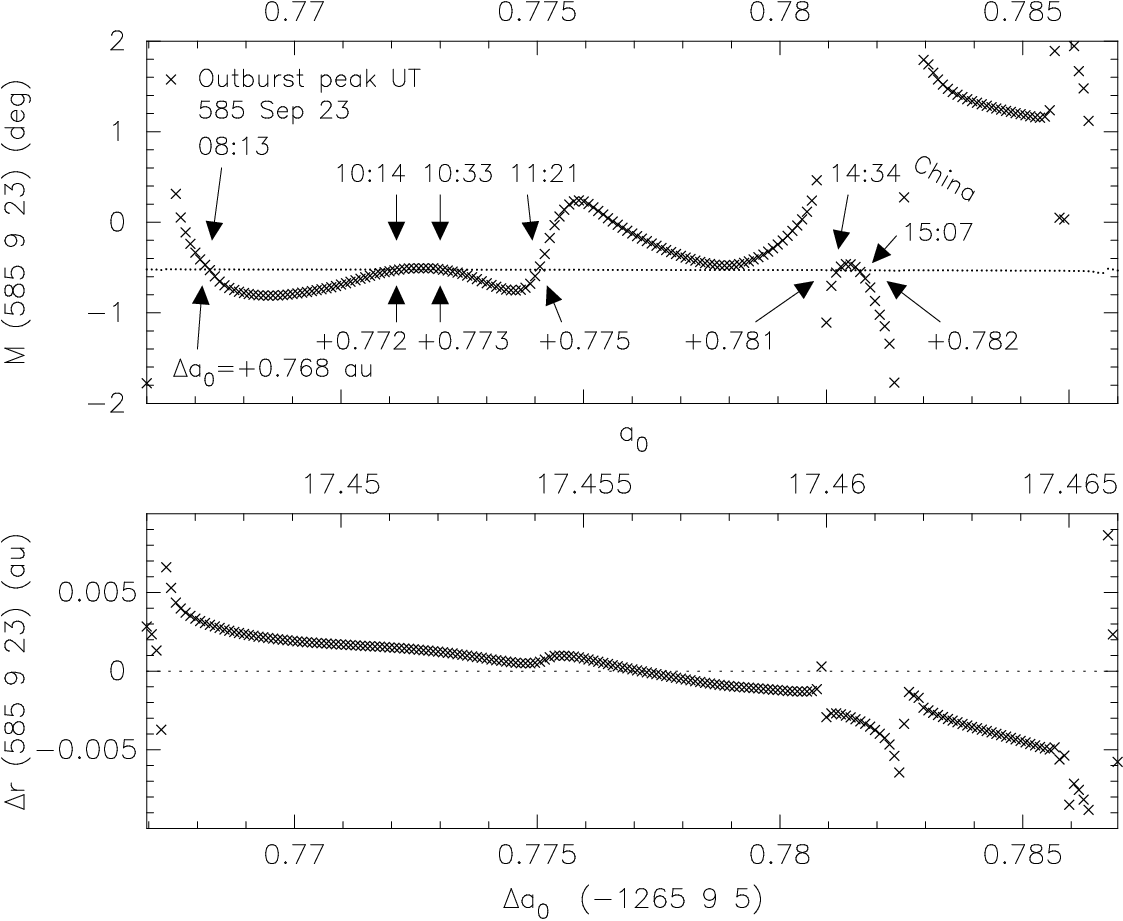}
\caption{In AD 585, on September 23, a tightly-packed stream of particles crossed the
Earth's orbital plane six separate times causing two outbursts visible to
China and likely three or four outbursts visible to the Maya.  In practice the outbursts from the six distinct trail encounters may overlap somewhat.
Particles with $M \approx -0.5\degr$ on this date are at Orionid node
when Earth is nearby.
}
\label{F585}
\end{figure}

Figure~\ref{F585} shows how this cluster of particles provides outbursts on
September 23 starting at 08:13 UT (02:13 AM local Maya time) and ending at
15:07 UT (10:07 PM local Daxing, China time), the first four outbursts
occurring when the Orionid radiant is above the horizon during hours of
darkness for the Maya (the fourth occurs 25 minutes before sunrise at 11:21),
while the last two occurred at 14:34 and 15:07 UT, during hours of darkness
in China.  The fact that the authors' results coincided with the actual
observation from China, and that remarkably both outbursts originated from
nearby resonant segments of the same trail, virtually assures the highest of
probabilities regarding the Maya outburst as well as confirming the China
observations.

The description of the outburst recorded at China reads: `several hundred
meteors scattered in all directions and came down.'  5th year of the Kaihuang
reign period of Emperor Gaozu of the Sui Dynasty, 8th month, day wushen [45];
[Bei shi: Sui Wen di ji] ch. 11. [Sui shu: Gaozu ji] ch. 1.
\citep[][pp.\ 313, 650]{pxj08}

Comparing the parameters for the probable Maya outbursts versus the Chinese
(Table \ref{T585}), the outbursts likely seen in the Maya area would have
been much more intense than the Chinese display for two reasons--the
along trail density |\fma| was around ten times greater than the Chinese,
$\sim$0.5 vs $\sim$0.05, and the nominal miss distance |\dr| from the Earth
was less than half that of the Chinese, 0.0012 vs 0.0027 au.

More Orionid outbursts from the 911 BC trail also occurred two nights later
on September 25 that would have been visible to the Maya
(Table~\ref{T585}). The particles
causing the two outbursts on this date were trapped in a 1:7 resonance, also
with Jupiter (Sec.~\ref{Smmr}).

Following the probable outburst on September 23, fifteen days later on
585 October 8 (Table \ref{TAevents}),
the ruler of Naranjo celebrated 40 years
of rulership.

In the specific results noted in the Tables and the above discussion, we quote nominal trail encounter times to the nearest minute UT, noting that the specific numbers allow comparison of results with other models, not that the numbers are actually accurate to the minute.
The integration results regarding the 585 China observations do, however,
indicate that Orionid trail encounters can be computed to an accuracy of
fractions of the day, i.e., correct observable longitude on Earth's
surface. Similarly, present epoch Orionid outbursts due to resonant trails 3
kyr old have been successfully computed \citep{sw07,sa14}.

\section{Mean motion resonance}
\label{Smmr}

\subsection{Orbital ratios}

Although trails only a few revolutions in age are generally compact enough to
produce meteor outbursts, the meteoroids causing the 417 and 585 outbursts
were rather older, $\sim$20 revolutions. Here (Sec.~\ref{Smmr}) we shall see
that these meteoroids remained synchronized in an integral orbit-to-orbit
ratio with Jupiter, defined as resonance. If Jupiter (or another planet) has
a strong gravitational force, particles can stay locked in such a synchrony
for a long time, up to thousands of years, and moreover the continuous action
of the resonance can allow a cluster of particles to remain compact
\citep{eb96, emel01, jenn02, sa13, sekhar16}. Indeed the salient point of
this work unequivocably demonstrates that sharply focused outbursts can
occur, in 417 and 585, from these very long-term resonant particles ejected
over 1000 years earlier.

\begin{figure}
\includegraphics[width=\columnwidth]{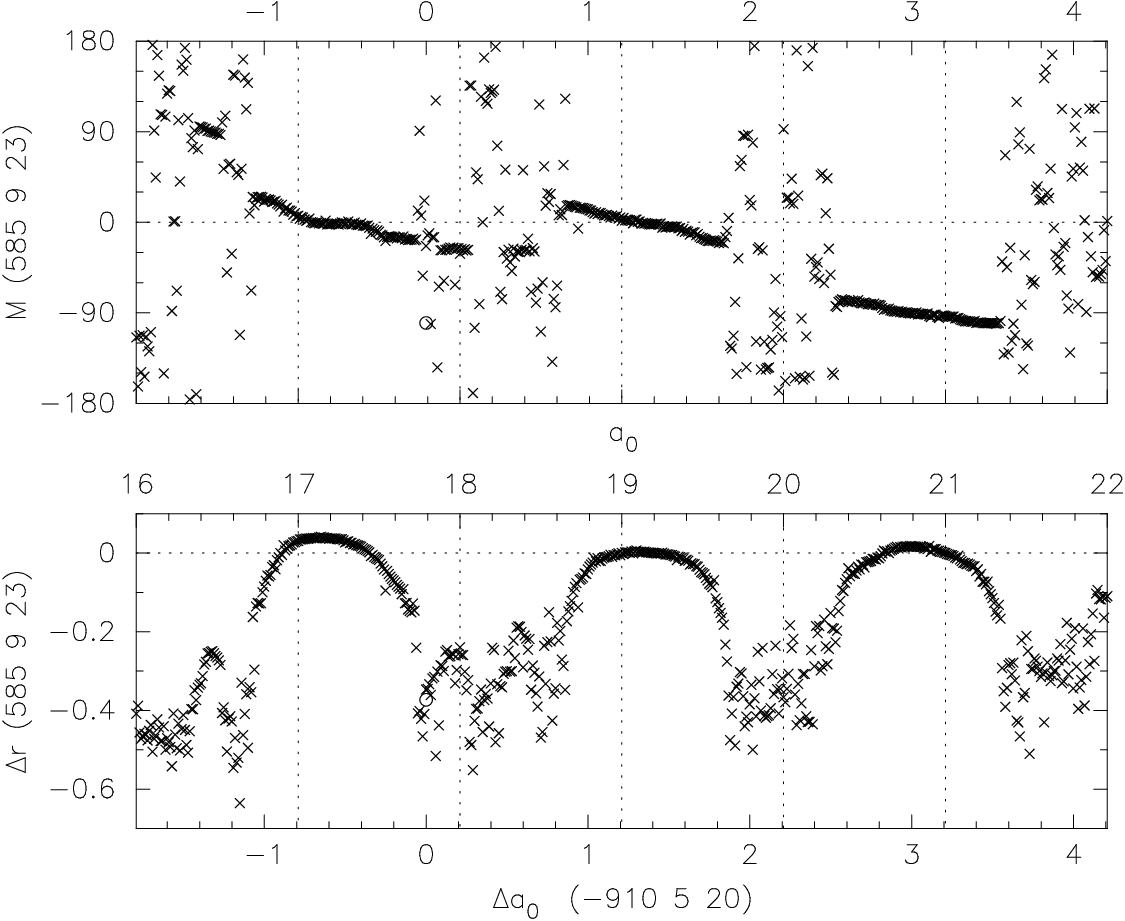}
\caption{
  Particles in AD 585 ejected by 1P/Halley in 911 BC over semi-major axis range $a_0$ = 16 to 22 au. Three major segments of particles are trapped in Jovian
  resonances (left to right) 1:6, 1:7, and 1:8. Upper panel indicates mean
  anomaly $M$ (degrees) in 585 September; ascending node is only
  $\sim$0.5\degr\ from perihelion, thus $M$ close to zero is necessary for an
  Orionid outburst in 585. Lower panel indicates radial distance (au) from
  Sun, expressed as $\dr = r_E - r_A$, thus \dr=0 coincides with Earth.
  Halley noted by small circle at $a_0 \approx$ 17.8 au.
}
\label{Ftrail}
\end{figure}

Figure~\ref{Ftrail} shows how a trail of particles, released from Halley during its perihelion passage in 911 BC, by the year 585 has broken up into three main segments of particles with a few smaller segments corresponding to weaker resonances as well as other individual particles scattered chaotically.
A plot of the same particles but shown in the year 417 would be similar except the groupings of particles would be shifted vertically.  
The three main segments indicate particles trapped tightly in resonances with the orbit of Jupiter in the ratio of 1:6, 1:7 and 1:8, left to right respectively.

In a rather rare distribution, both the 1:6 and 1:7 orbital ratios indicate large groupings of particles simultaneously at $M \approx 0$ and $\dr \approx 0$, setting up potentially very strong outbursts, given the appropriate time of night for visual observation. As we indicated in Section \ref{Sres}, this is indeed the case
for the 1:7 resonance (i.e., particles exist with small enough \dr; for the
1:6, on closer examination than can be seen in Fig.~\ref{Ftrail}, particles
with exactly suitable $M$ turn out to be a little too distant in \dr).
The 1:6 and 1:7 coincidence occurs at this time because AD 585 is close to 126
Jovian periods $P_J$ after 911 BC: both 6 and 7 divide 126.  Resonant 1:8
particles are $\sim$$\frac{1}{4}$ revolution ($\sim$90\degr\ in $M$) behind.

In actuality, the resonant period differs from the exact integer ratio
multiple of $P_J$ \citep[][p.~331]{md99} \citep[][pp.~54--55]{sa14}.
\citet[][table 1]{emel01} computes semi-major axes of resonances in several
streams and these differ from the nominal resonant semi-major axes $a_n$
which would come from the $P_J$ multiple and Kepler's 3rd Law. However, the
difference is slight and for explanatory purposes we neglect it, thus 18
periods of 1:7 = 126$P_J$ etc.

\subsection{Resonant argument and verification of resonance}

\begin{figure}
\includegraphics[width=\columnwidth]{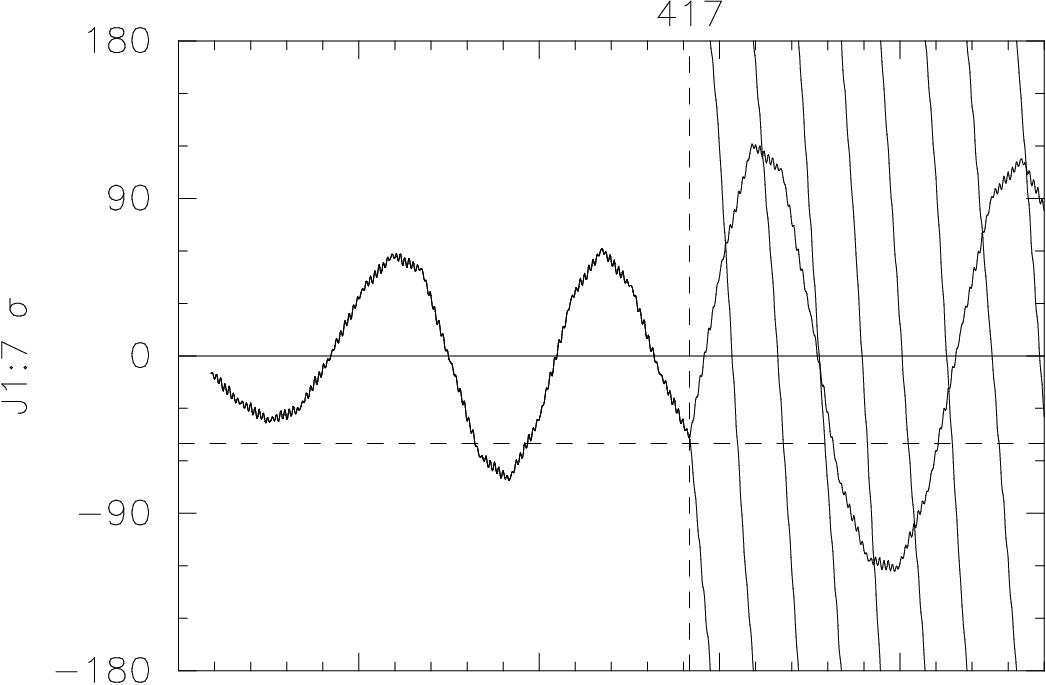}
\protect\\[2ex]
\includegraphics[width=\columnwidth]{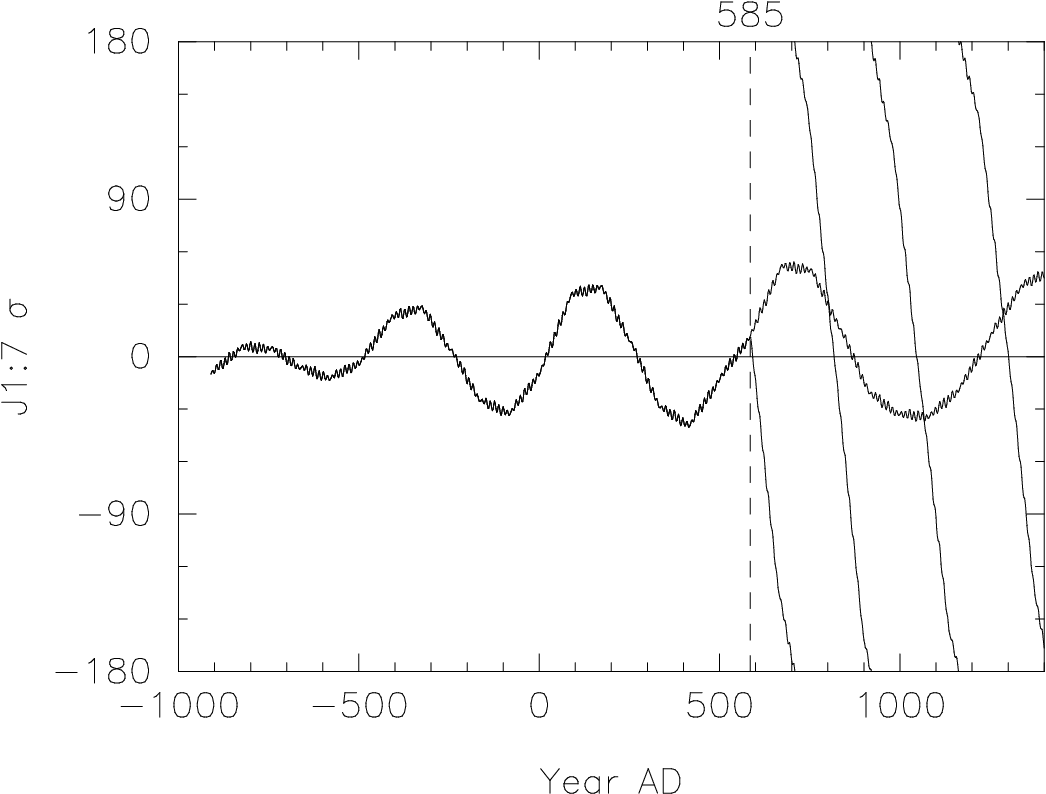}
\caption{Four particles ejected into the Jovian 1:7 resonance at 1P/Halley's
  911 BC perihelion passage.  In the upper panel two of these particles, with $\da \approx +1.180$
  (Table \ref{T417}), virtually coincide until 417 when one particle leaves resonance upon approach to Earth.  In the lower panel two particles, with $\da \approx +1.343$
  (Table \ref{T585}) are coincidental until 585 when one leaves resonance upon approach to Earth. One particle in each panel continues in resonance indicated by librations within the $\pm 180\degr$ boundaries.
  The libration centre is $\sigma \approx 0\degr$.
  Particles out of resonance are indicated by roughly parallel lines extending the full range of $-180\degr$ to $+180\degr$.
  All four particles show librating resonant argument $\sigma$ over the whole
  interval from ejection till Earth approach, as do all other particles from
  911 BC contributing to outbursts in both years.
  The particles in 417 (upper panel) are at $\sigma \approx -50\degr$
  when they approach Earth, somewhat ahead of the libration centre.
  }
\label{Fsigma911bc}
\end{figure}

Whereas Fig.~\ref{Ftrail} shows many particles, resonant groupings in various
$a_0$ ranges being evident, resonant behaviour is verified by plotting the
\textit{resonant argument}, denoted by $\sigma$, for individual particles
(Fig.~\ref{Fsigma911bc}).

For a 1:7 resonance $\sigma$ is defined as:
\begin{equation}
   \sigma = \lambda_J - 7\lambda + 6\varpi
\label{eq:1:7}
\end{equation}
and for a 1:6 resonance (not shown in Fig.~\ref{Fsigma911bc}), 
\begin{equation}
   \sigma = \lambda_J  - 6\lambda + 5\varpi
\label{eq:1:6}
\end{equation}
where $J$ is Jupiter and $\lambda$, the \textit{mean longitude}, is defined by:
\[
   \lambda = \varpi + M
\]
and $\varpi$ is the \textit{longitude of perihelion}.   We define $\varpi$ in relation to a retrograde orbit, 
\begin{equation}
   \varpi = \Omega - \omega
\label{eq:lop}
\end{equation}
where $\Omega$ is longitude of ascending node and $\omega$ is argument of perihelion. In the case of a retrograde orbit, $\omega$ is subtracted from
$\Omega$ \citep{ws93,sa14} instead of being added as in the conventional
definition of $\varpi$ for a prograde orbit.
Equations (\ref{eq:1:7}) and (\ref{eq:1:6}) for the resonant arguments
$\sigma$ hold because the d'Alembert rules are satisfied
\citep[][p.~250]{md99} \citep[][p.~53]{sa14}. The slowly varying terms
($\lambda_J$ and $\lambda$ being rapidly varying) are chosen to be single
terms in $\varpi$ because with the particles having high $e$, these
resonances are of the eccentricity type as described in
\citet[][sec.~2]{peale76}.

Resonant behaviour is identified by verifying that $\sigma$ librates, or
oscillates back and forward within the confines of a single
\textit{libration zone\/} \citep{emel01} or
\textit{resonant zone\/} \citep[][sec.\ 4.20]{ka17}.
The above choice of definitions
(eq.\ \ref{eq:1:7}, \ref{eq:1:6}, \ref{eq:lop}) results, for these 1:6 and
1:7 resonances, in particles librating around the centre of a resonant zone
at $\sigma=0$, the zone's boundaries being at $\sigma = \pm180\degr$ (see
Fig.~\ref{Fsigma911bc}).

A 1:$n$ resonance divides the 360\degr\ mean longitude of the orbit into $n$
resonant zones \citep{emel01}. So for the 1:7, there are seven resonant
zones, yielding a length of $\sim$52\degr\ per zone and the centre
of each zone would then occur at the centre of that 52\degr\ interval (i.e.,
$\pm180\degr/7 = \pm26\degr$ in $\lambda$). Here $\sigma$ effectively
amplifies, by 7 times, the longitude's displacement relative to the resonance
centre (cf.\ eq.\ \ref{eq:1:7}).

Analogously to the 1:7, the 1:6 orbital ratio produces six zones around the
360\degr\ of the orbit, any resonant particle librating within one zone whose
total extent is $\pm$30\degr\ in $M$ about the respective resonance centre.
In the 1:6 or 1:7, each zone of the respectively six or seven zones
corresponds to one revolution of Jupiter ($\sim$12 yr), i.e.\ the extent of
each zone equalling $\pm \frac{1}{2} P_J$.

\subsection{Particles trapped in 1:7 resonance: examples}
\label{S1:7ex}

Figure~\ref{Fsigma911bc} examines the dynamics of this resonant trapping in detail, i.e., the libration vs circulation of the resonant argument $\sigma$ corresponding to being trapped in resonance or not.
Interestingly enough, even though all four particles are ejected in 911 BC
into a 1:7 resonance with Jupiter, two particles closely approach Earth in
417, corresponding to an outburst that year, while the other two particles
cause a second outburst in 585.
Upper and lower panels show all four particles undergoing resonant librations
until 417 and 585 respectively.
Whereas most of the particles stay in resonance, indicated by the continuous oscillations, some particles making a close passage to Earth are thrown out of resonance, illustrated in Fig.~\ref{Fsigma911bc} by a particle exceeding $-$180\degr\ in both 417 and 585.

\begin{table}
\centering
\caption{Mean motion resonances, Jovian 1:6 and 1:7, causing observable and
  likely observable outbursts: $a_n$ nominal resonance location
  \citep[][sec.\ 8.4]{md99};
  $a_0$ osculating semi-major axis at ejection; Diff =
  difference in days with recorded event, where + indicates the event
  followed the probable outburst by the given number of days.}
\label{Tmmr}
\begin{tabular}{@{}cc@{\,\,\,}c@{\,\,\,}cccc@{}}
\hline
Year       & Trail  & Reson & $a_n$ &  $a_0$ & Obs.  & Diff \\
\hline
417 Sep 24 & \0911 BC & 1:7 & 19.03 & 18.966 & Maya  & +5 \\
           & \0911 BC & 1:7 & 19.03 & 18.971 & Maya  & +5 \\
           & \0911 BC & 1:7 & 19.03 & 
           18.974 & Maya  & +5 \\
585 Sep 23 &  1266 BC & 1:6 & 17.17 & 17.447 & Maya  & +15 \\
           &  1266 BC & 1:6 & 17.17 & 17.451 & Maya  & +15 \\
           &  1266 BC & 1:6 & 17.17 & 17.452 & Maya  & +15 \\
           &  1266 BC & 1:6 & 17.17 & 17.454 & Maya  & +15 \\
           &  1266 BC & 1:6 & 17.17 & 17.460 & China & 0 \\
           &  1266 BC & 1:6 & 17.17 & 17.461 & China & 0 \\
585 Sep 25 & \0911 BC & 1:7 & 19.03 & 19.238 & Maya  & +13 \\
           & \0911 BC & 1:7 & 19.03 & 19.134 & Maya  & +13 \\
\hline
\end{tabular}
\end{table}

Table~\ref{Tmmr} shows the 1:7 resonant returns from the same 911 BC trail in 417 and 585. 
The osculating semi-major axis $a_0$ at ejection is smaller for AD 417, $\approx$ 18.97 au and larger for AD 585, $\approx$ 19.2 au.  Thus two separate initial sizes
of the semi-major axis $\mathit{a_0}$, corresponding to the two separate
years 417 and 585 when the particles reach Earth, were drawn into the same
range of resonant librations about the nominal resonant $a_n$ = 19.03 au.
The values 18.97 and 19.2 both fall within the quite extensive range of $a_0$
under the influence of 1:7 (see the main central concentration in
Fig.~\ref{Ftrail}).

In fact, Figure~\ref{Fsigma911bc} shows when the particle is at the front or back
of its oscillation as it repeatedly rebounds between
$\pm180\degr$; the resonance centre is $\sigma \approx 0\degr$ in this case.  The sign of $\sigma$ is arbitrary but with our definition (Eq.\ \ref{eq:1:7})
positive $\sigma$ corresponds to particles behind centre and negative $\sigma$ corresponds to particles ahead of the resonance centre. 
The initial negative slope shown in the upper panel (Fig.~\ref{Fsigma911bc})
shows forward drift initially where $a<a_n$
while the initial positive slope in the lower panel indicates rearward drift initially where $a>a_n$.

\citet{rendtel07} and \citet{arlt08} suggest enhanced Orionid activity can
reoccur for a few consecutive years. Particles undergoing 1:6 or 1:7 resonant
librations can periodically be a few years ahead or behind while remaining in
a single resonant zone.  Figure \ref{Fsigma911bc} shows a \da=+1.180 particle
$\sim$1.6 yr forward of the resonance centre in 417 (upper panel:
1.6 yr ahead of centre is $\sigma \approx -50\degr$ since
360\degr\ in $\sigma$ corresponds to $P_J \approx 11.9$~yr) and a \da=+1.343
particle near the centre in 585 (lower panel).  Strictly speaking,
regarding the continuous resonance from 911 BC until AD 417, the resonance
centre does not reach Earth until 419, which is 112 (= 16 $\times$ 7) $P_J$
after 911 BC. Thus precisely, 419 and 585 are separated by 2 resonant
revolutions = 14$P_J$. In practice, however, 417 and 419 are within the same
resonant zone, particles in 417 September
reaching the ascending node $\sim$1.6 yr ($\sim$50\degr\ in $\sigma$)
before those at the resonance centre do so.\footnote{
It is not the purpose of this paper to present details of the locations of
resonant zones, such as listing perihelion passage times of the centres of
these zones, but as we refer to these times in a few cases, we note
that they can be computed by constructing diagrams as in
\citet[][p.54]{sa14}. Sekhar has made corresponding computations for resonances
in other streams \citep{sekhar14}, resonances with other planets \citep{sa13}
and three body resonances \citep{sekhar16}.
}

\subsection{The year AD 585, the 1266 BC trail and returns from 1:6 resonance}

The 585 outbursts on September 23rd were due to a 1:6 Jovian resonance of
particles released at the 1266 BC passage of 1P/Halley (librating $\sigma$
plots analogous to Fig.~\ref{Fsigma911bc} confirmed but not shown here).
Table~\ref{T585} gives the parameters of the 585 outbursts and
Table~\ref{Tmmr} shows the details of the mean motion resonance. As noted in
Section \ref{Sres}, the outbursts from the 1266 BC trail with $a_0 \approx
17.45$ au, where the nominal $a_n = 17.17$ au, were visible both to the Maya
and the Chinese. Contrary to the earlier situation describing the 417
outburst $\sim$1.6 yr
prior to the 419 resonance centre (Sec.~\ref{S1:7ex}),
AD 585 was right on the 1:6 resonance centre from the 1266 BC ejection, i.e.,
26 revolutions at 1:6 (156 $P_J$).

Since Comet Halley was 1:6 resonant from 1404 BC to 690 BC
\citep[][p.~53]{sa14}, any particles ejected in that interval have an
increased chance of being locked in the same 1:6 resonance.  In fact, typical meteoroid ejection speeds are large enough to bring other resonances
like 1:7 within range too (cf.\ Fig.~\ref{Ftrail}).
Moreover during that interval the returns when Halley is nearest the 1:6
resonance centre are particularly favourable for populating 1:$n$ resonances;
911 BC is best, followed by 1266 BC, then 1198 BC, then 986 BC.
This is supported by computations of present epoch outbursts from trails
created at these favourable returns \citep[][table 1]{sw07}
\citep[][table 1]{sa14}.
In actuality, Comet Halley was very close to the 1:6 resonance centre at its 911 BC
return (particles starting only slightly below zero in
Fig.~\ref{Fsigma911bc}).

\section{Conclusions}

The authors have presented evidence that there was a high probability of
Orionid outbursts occurring in the years 417 and 585. Particles trapped in
resonance from Comet Halley's perihelion passages in 1266 BC and 911 BC would
have provided sharp, dense and well-defined displays under near ideal
conditions for the Maya.
The authors confirmed the Chinese observation in 585, verified by particles
from the 1266 BC Comet Halley trail resonant with Jupiter in a 1:6 orbital
ratio. Further, in a rather unusual display of orbital
gymnastics, this same tightly packed stream segment that caused
the China outburst would only hours earlier have caused a dramatic display
visible to Maya observers. Although the Orionids have been an annual meteor
shower since at least AD 288 \citep{pxj08}(p. 309), \citep{zhuang77}(entry 100, p. 205), \citep{hasegawa93}(entry 160, p. 215), actual outbursts
prior to the year 900 seem to have been a rarity as evidenced by only two
recorded observations by China.

\section*{Acknowledgements}

Plots used Prof.\ Timothy J. Pearson's excellent \textsc{pgplot} subroutine
library. Astronomy at Armagh is supported by the N. Ireland Dept.\ for
Communities.  The authors are very grateful to the editor and reviewers; their comments were responsible for a significant improvement in our paper.



\bibliographystyle{mnras}
\bibliography{orionid-maya}


\appendix

\section{Data base of Maya Classic Period dates integrated for possible
Orionid outbursts}

\begin{table*}
\centering
\caption{
  ORI Data Set. Columns: Julian Calendar Date \citep{ms12}, J2000 Solar Longitude;
  location, site where the date is inscribed. *1 refers to a portable object
  \citep[see][]{mm38}. *27 refers to one of four extant painted screeenfold
  books. As a matter of note, one event known as a `Star' War was recorded on
  692 Oct 5, within the possible time frame of the Orionids. `Star War'
  refers to a war event that is denoted by a hieroglyph containing the sign
  for `star,' known as the `star-over-earth' hieroglyph.  In his publication
  about the `Star War' hieroglyph, \citet{aldana05} proposes a connection
  between sporadic meteors and the Star War event through purely historical
  reassessment and a `coordinate free' environment (i.e., no correlation
  constant) and therefore unfortunately no correlations to specific dates can
  be determined; he discounts meteor showers because `Modern science's
  ability to calculate this variation becomes increasingly suspect the
  farther it projects into the past, making reconstructions for the Classic
  period unreliable' \citep[p.314]{aldana05}. Sporadic meteors are those
  meteors which are unrelated to established meteor showers. Although it is
  possible that `star-over-earth' does refer to sporadic meteors, it is
  nearly impossible to prove at this juncture and furthermore sporadic meteors
  are very common \citep{wiegert09}. Meteor outburst predictions however,
  have proven to be very accurate using computerized orbital integrations
  \citep{arlt99,sato17,egal19,vaubaillon19} both with recent orbital
  revolutions and in particular regarding resonances produced over centuries
  \citep{ab99,ae02,rendtel07,sw07,trigo07}. The correlation
  constant of 584286 (actually the base date of the Maya calendar in terms of
  the Julian Day Number) is a virtual certainty using the \citet{ms12}
  argument and the most recent high precision radiocarbon dating
  \citep{kennettetal13}. Using this correlation constant the authors produce
  the specific dates in this table, however, only one out of 30 associated
  events contained the `Star War' hieroglyph, a date with a year (692) in which the
  authors found no Orionid outburst.
}
\label{TAevents}
\begin{tabular}{rccc}
\hline
   & Date Julian & Solar Long & Location \\ \hline
   
*1 & 320 Sep 17  &  198\fdg445  & Artifact* (coastal Guatemala)  \\
2 & 378 Oct \07 &  217\fdg531  &  Yaxchilan, Chiapas, Mexico  \\
3 & 402 Sep 28  &  208\fdg369  &
Yaxchilan, Chiapas, Mexico  \\
4 & 417 Sep 29  &  209\fdg519  &
Rio Azul, Peten, Guatemala  \\
5 & 508 Sep 25 & 205\fdg196  &
Palenque, Chiapas, Mexico  \\
6 & 524 Sep 19 &  199\fdg126  &
Palenque, Chiapas, Mexico  \\
7 & 557 Sep 16 & 195\fdg694  &
Tikal, Peten, Guatemala \\
8 & 585 Oct \08 & 217\fdg435  &
Naranjo, Peten, Guatemala \\
9 & 614 Sep 23 & 202\fdg022  &
Bonampak, Chiapas, Mexico  \\
10 & 625 Sep 30 &  209\fdg171  &
La Corona, Peten, Guatemala  \\
11 & 629 Oct \02 &  211\fdg147  &
Caracol, Belize  \\
12 & 630 Oct \02 &  210\fdg885  &
Naranjo, Peten, Guatemala  \\
13 & 634 Sep 28 & 206\fdg874  &
Caracol, Belize  \\
14 & 638 Sep 20 &  198\fdg892  &
Caracol, Belize  \\
15 & 640 Sep 25 &  204\fdg347  &
Caracol, Belize  \\
16 & 649 Sep 25 &  204\fdg035  &
Palenque, Chiapas, Mexico  \\
17 & 664 Oct \01 & 210\fdg176  &
La Corona, Peten, Guatemala  \\
18 & 667 Sep 20 &  198\fdg456  &
La Corona, Peten, Guatemala  \\
19 & 675 Sep 22 &  200\fdg392  &
La Corona, Peten, Guatemala  \\
20 & 689 Sep 26 &  204\fdg774  &
La Corona, Peten, Guatemala  \\
21 & 692 Oct \05 & 213\fdg983  &
Tonina, Chiapas, Mexico  \\
22 & 714 Sep 24 &  202\fdg380  &
Naranjo, Peten, Guatemala  \\
23 & 724 Sep 24 &  202\fdg810  &
Palenque, Chiapas, Mexico  \\
24 & 735 Sep 23 &  201\fdg005  &
Tikal, Peten, Guatemala  \\
25 & 767 Sep 24 &  201\fdg790  &
Itzan, Peten, Guatemala  \\
26 & 773 Sep 23 &  201\fdg254  &
Copan, (western) Honduras  \\
*27 & 775 Sep 30 &  207\fdg711  &
Dresden Codex*  \\
28 & 836 Sep 24 &  202\fdg094  &
Hecelchakan, Campeche, Mexico  \\
29 & 859 Oct \06 & 213\fdg162  &
Caracol, Belize  \\
30 & 874 Sep 28 &  205\fdg334  &
Seibal, Peten, Guatemala  \\

\hline
\end{tabular}
\end{table*}


\bsp	
\label{lastpage}
\end{document}